\documentclass[nofootinbib,amsmath,twocolumn,notitlepage,preprintnumbers]{revtex4-1}
\usepackage{multirow}
\usepackage{amssymb,esvect,amsmath,graphicx,latexsym,amsthm,slashed,eso-pic}
\usepackage{hyperref}
 \DeclareMathOperator{\kev}{keV}  \DeclareMathOperator{\gev}{GeV}    \DeclareMathOperator{\cm}{cm}           
         \newcommand{\cO}{{\cal O}}

 \def\d{{\rm d}}   
\newcommand{\pL}{\left(} \newcommand{\pR}{\right)} \newcommand{\bL}{\left[} \newcommand{\bR}{\right]} \newcommand{\cbL}{\left\{} \newcommand{\cbR}{\right\}}  
\newcommand{\beq}{\begin{equation}} \newcommand{\eeq}{\end{equation}}
\newcommand{\bea}{\begin{eqnarray}} \newcommand{\eea}{\end{eqnarray}}

\newcommand{\vev}[1]{\langle {#1} \rangle}
\newcommand{\tenx}[1]{\times 10^{#1}}

\def\lsim{\mathrel{\raise.3ex\hbox{$<$\kern-.75em\lower1ex\hbox{$\sim$}}}}
\def\gsim{\mathrel{\raise.3ex\hbox{$>$\kern-.75em\lower1ex\hbox{$\sim$}}}}

\newcommand{\Eq}[1]{Eq.~(\ref{#1})}  
\newcommand{\Sec}[1]{Sec.~\ref{#1}}  
\newcommand{\Fig}[1]{Fig.~\ref{#1}}

\begin{document}

\preprint{FERMILAB-PUB-18-032-A}

\title{Robust Constraints and Novel Gamma-Ray Signatures of Dark Matter That Interacts Strongly With Nucleons}
\author{Dan Hooper$^{a,b,c}$ and Samuel D.~McDermott$^a$}
\affiliation{$^a$Fermi National Accelerator Laboratory, Center for Particle Astrophysics, Batavia, IL, USA}
\affiliation{$^b$University of Chicago, Kavli Institute for Cosmological Physics, Chicago IL, USA}
\affiliation{$^c$University of Chicago, Department of Astronomy and Astrophysics, Chicago IL, USA}

\date{\today}

\begin{abstract}

Due to shielding, direct detection experiments are in some cases insensitive to dark matter candidates with very large scattering cross sections with nucleons. In this paper, we revisit this class of models, and derive a simple analytic criterion for conservative but robust direct detection limits. While large spin-independent cross sections seem to be ruled out, we identify potentially viable parameter space for dark matter with a spin-dependent cross section with nucleons in the range of $10^{-27} {\rm cm}^2  \lsim \sigma_{{\rm DM}-p} \lsim10^{-24} \, {\rm cm}^{2}$. With these parameters, cosmic-ray scattering with dark matter in the extended halo of the Milky Way could generate a novel and distinctive gamma-ray signal at high galactic latitudes. Such a signal could be observable by Fermi or future space-based gamma-ray telescopes. 
\end{abstract}

\maketitle

\section{Introduction}

If the particles that make up the dark matter of our universe interact strongly with the Standard Model, such interactions would be expected to generate large event rates in direct detection experiments, assuming that the dark matter is able to reach the detectors with a standard velocity distribution. If the cross section for these interactions is very large, however, direct detection experiments can be effectively shielded by the Earth or its atmosphere~\cite{Kavanagh:2017cru} (see also Refs.~\cite{Goodman:1984dc,Starkman:1990nj, Collar:1992qc, Collar:1993ss, Hasenbalg:1997hs, Foot:2011fh, Kouvaris:2014lpa, Kouvaris:2015laa, Foot:2014osa, Clarke:2015gqw, Bernabei:2015nia, Emken:2017erx, Davis:2017noy, Mahdawi:2017cxz}). In this way, such shielding could render dark matter with large scattering cross sections invisible at direct detection experiments. For these reasons, qualitatively different theoretical and experimental considerations are necessary when considering dark matter candidates with very strong interactions with the Standard Model.

If the dark matter can annihilate in the present epoch, the observed heat flow of the Earth can be used to provide a very strong constraint on the dark matter's elastic scattering cross section with nuclei, largely insensitive to this shielding loophole~\cite{Mack:2007xj}. Non-annihilating dark matter can be broken into a few further subcategories. Scalar dark matter that does not annihilate with itself can accumulate in the center of compact astrophysical objects and eventually exceed the Chandrasekhar limit. The observation of old neutron stars provides very stringent constraints on this type of model~\cite{Kouvaris:2010jy, McDermott:2011jp, Bramante:2013hn}. Similar bounds also apply to dark matter in the form of a very massive fermion with large interactions with both itself and nucleons~\cite{Kouvaris:2011gb, Bramante:2013nma}.

Constraints on light, non-annihilating fermionic dark matter with very large cross sections with baryons are less firmly established. One particularly interesting example of this class of models is a bound state of $uuddss$ quarks, which may have formed in the early universe as a byproduct of the Standard Model baryon asymmetry~\cite{Zaharijas:2004jv}. It has been argued that such a six quark configuration -- known as the $H$ dibaryon~\cite{Jaffe:1976yi} or the $S$ sexaquark~\cite{Farrar:2017eqq} -- is a bound state of QCD. Although lattice simulations support the existence of a weakly bound dibaryon with mass just below twice the mass of the $\Lambda$ baryon~\cite{Beane:2010hg, Beane:2012vq}, a more tightly bound and thus stable or cosmologically metastable six-quark state cannot be ruled out. In particular, a very deeply bound dibaryon could have a small enough overlap with lattice sources to have evaded notice~\cite{SavageBeaneCommunication}. In order for a dibaryon to be the dark matter, cosmological metastability is a necessary but not sufficient condition. The dibaryon must also have a hadronization rate at the time of the QCD phase transition to have been produced in greater abundance than standard baryons. If both of these conditions are met, dibaryon dark matter would evade the constraints described in the previous paragraphs.

In this paper, we revisit the constraints on non-annihilating dark matter with a large scattering cross section with nucleons, and discuss the astrophysical gamma-ray signatures in this class of models. We find that direct detection constraints exclude the entirety of the parameter space in the case of dark matter with spin-independent interactions, for all masses above a few hundred MeV. In contrast, there is an open region of parameter space in which the dark matter could possess a very large spin-dependent coupling to nucleons. In exploring the gamma-ray signatures associated with this class of models, we find that the scattering of cosmic-ray protons with dark matter in the extended halo of the Milky Way could lead to a potentially observable signal at high latitudes, with distinctive spectral and morphological characteristics.

The remainder of this paper is organized as follows. In~\Sec{BBN-and-CMB}, we discuss and summarize prior work on early universe effects of dark matter with a large scattering cross section with nucleons, in particular in regards to Big Bang nucleosynthesis (BBN) and the cosmic microwave background (CMB). In~\Sec{DD-scatt}, we describe a simple method of deriving conservative and robust bounds on the dark matter scattering cross section with nucleons. In~\Sec{gamma-ray-sec}, we discuss the gamma-ray signatures that can arise within this class of models. Throughout this study, we will refer to dark matter as $\psi$ and we will assume it does not annihilate in the present epoch.


\section{Constraints on Dark Matter With Large Scattering Cross Sections With Nuclei}

We will order our discussion chronologically with respect to cosmic time, focusing on the impact on the light element abundances during BBN, the power spectrum of perturbations at the formation of the CMB, and on direct or indirect detection experiments.

\subsection{BBN and CMB}
\label{BBN-and-CMB}

At early times, the high-temperature photon bath readily dissociates nuclear bound states. This continues until the temperature is low enough that photons of sufficient energy to unbind the states are Boltzmann suppressed by more than the baryon-to-photon ratio, $\eta_B = n_B/n_\gamma \simeq 6\tenx{-10}$, where $n_B=n_p+n_n$. The smallness of $\eta_B$ implies that the temperature must satisfy $B_A/T \gtrsim \ln(\eta_B^{-1}) \sim \cO(10)$ before the Boltzmann suppression of photons is severe enough that a typical nucleus of binding energy $B_A$ is likely to survive or to continue synthesizing heavier nuclei. Standard BBN commences when deuteron survival becomes likely, at temperatures around $T_\d \sim B_\d/\ln(\eta_B^{-1}) \sim 100\kev$.

Considering the dissociation of nuclei by dark matter, for the range of cross sections of interest here, the dark matter and Standard Model material have the same temperature during BBN and thus collide with the same kinematics.  Given the small error bars on measurements of the primordial deuterium and helium abundances, we set the rough requirement to affect the development of the nucleosynthetic chain by $n_\psi \sigma_{\psi \d} v_\psi \gtrsim n_\gamma \sigma_A v_p$, where $\sigma_A \simeq 10^{-26} \cm^2$ is the Thomson cross section and $v_i$ is the velocity of a particle of type $i$. Plugging in the value of $\eta_B$, we require approximately $\sigma_{\psi p}/m_\psi^{3/2} \gtrsim 3\tenx{-19} \cm^2/\gev^{3/2}$ to impact BBN. We are able to verify this at better than the order of magnitude level using a modified version of the {\tt AlterBBN} code~\cite{Arbey:2011nf, abbn}. In brief, and in rough agreement with Ref.~\cite{Cyburt:2002uw}, we find that the smallness of $\eta_B$ indicates that $\sigma_{\psi p}$ must be quite large to have any observable effects during the BBN epoch.

At later times, dark matter will impact the power spectrum of the CMB. The most recent bounds on this effect have been derived by Gluscevic and Boddy~\cite{Gluscevic:2017ywp}.\footnote{Although these CMB constraints are somewhat less stringent than those presented in Ref.~\cite{Dvorkin:2013cea}, the results of the latter study rely on a linear scaling relationship which applies only for $m_\psi \gg m_p$. The constraints of Ref.~\cite{Gluscevic:2017ywp}, in contrast, are applicable for all dark matter masses. We also note that the constraints based on the stability of the Milky Way's disk~\cite{Mack:2007xj} are less stringent than those in Ref.~\cite{Gluscevic:2017ywp}.}  For large cross sections, the dark matter exerts a drag force on Standard Model matter which affects the shape and the amplitude of the high-$\ell$ CMB power spectrum to a degree that is ruled out by current measurements. This bound has no upper value: for the CMB power spectrum, there is no analogue of the shielding effect for direct detection experiments that we are about to discuss.


\subsection{Direct Detection}
\label{DD-scatt}

In the present era, dark matter from our own galaxy may scatter off of low-threshold detectors. The lack of observed scattering in direct detection experiments has been used to rule out dark matter with a weak-scale scattering cross section with nucleons. It is well-known that the greatest rate can be achieved if the dark matter has spin-independent couplings to Standard Model particles~\cite{Goodman:1984dc}. This can be seen from comparing the spin-independent scattering cross section (which gives the well-known $A^2$ coherent enhancement) to the spin-dependent cross section, which does not provide coherent scattering (and thus does not necessarily lead to increased rates for larger nuclei):
\beq \label{sigT}
\sigma_{\psi A} = \sigma_{\psi p} \pL\frac{\mu_{\psi A}}{\mu_{\psi p}}\pR^2 \times \cbL \begin{array}{cc} \bL Z+\frac{f_n}{f_p}(A-Z)\bR^2 & {\rm (SI)} \\ \bL \vev{S_p} + \frac{f_n}{f_p} \vev{S_n} \bR^2 \frac43 \frac{J_A+1}{J_A} & {\rm (SD)} \end{array} \right..
\eeq
As a result, the experimental sensitivity to spin-dependent scattering is significantly reduced.

\begin{figure*}[htbp]
\begin{center}
\includegraphics[width=0.47\textwidth]{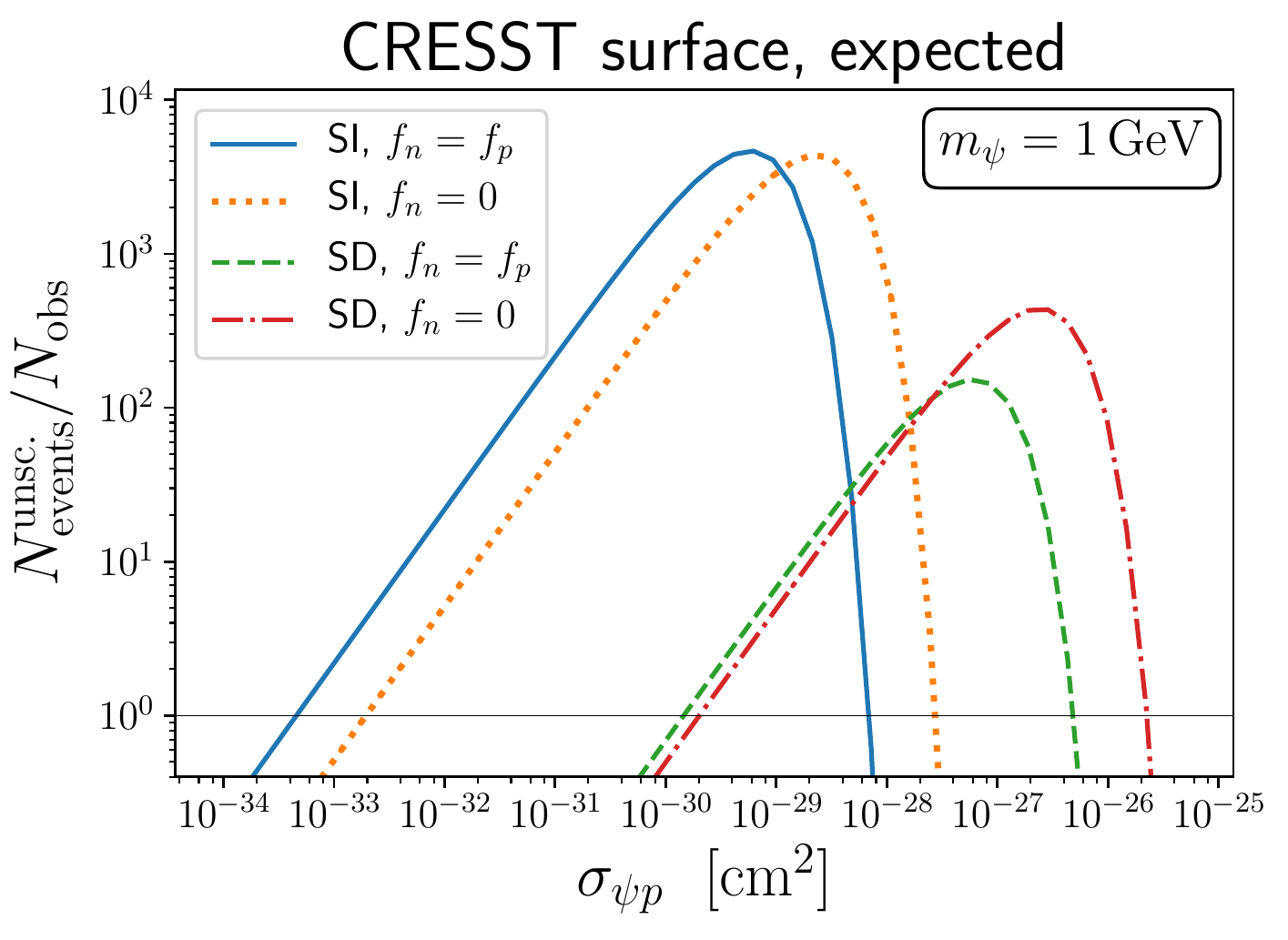}
 ~~~~\includegraphics[width=0.48\textwidth]{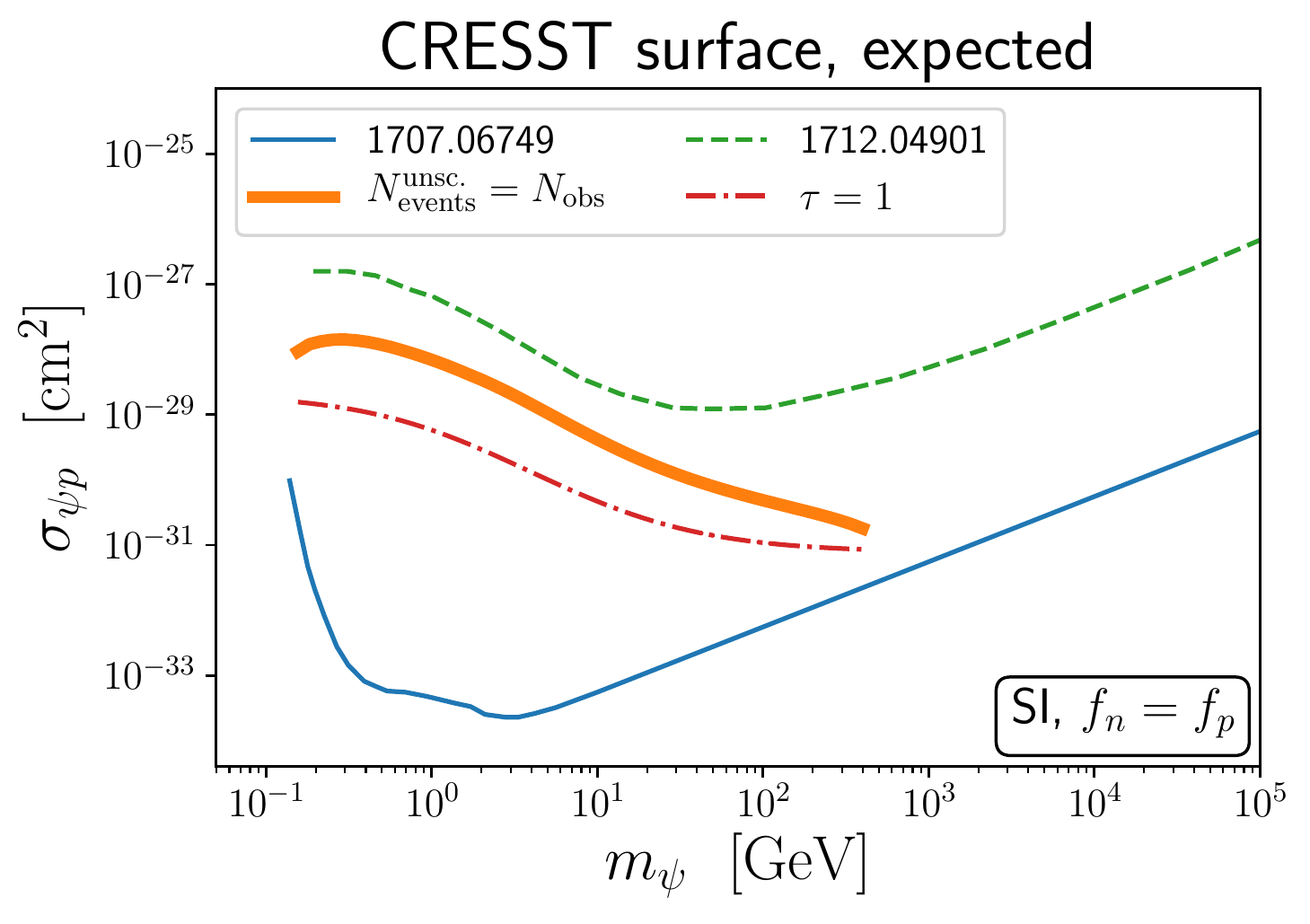}
\caption{In the left frame, we plot the number of unscattered events expected in the CRESST surface run~\cite{Angloher:2017sxg} as a function of $\sigma_{\psi p}$ for different couplings to nucleons, as in \Eq{n_expec}. For very large values of the cross section, the rate becomes exponentially suppressed due to scattering in the atmosphere. In the right frame, the lower solid line (labeled 1707.06749) denotes the limit as derived from the CRESST surface run~\cite{Angloher:2017sxg}; for values of the cross sections near this curve, the dark matter is unlikely to undergo any scattering with the shielding. Along the dot-dashed red line, the number of unscattered events is maximized as in \Eq{sig-max} ($\tau=1$). The thick orange line indicates where the number of unscattered events equals the number of observed events. We can conservatively and robustly rule out the range of cross sections between the two solid contours. We also show as a dashed green line the upper boundary of this region as derived in Ref.~\cite{Kavanagh:2017cru, Davis:2017noy}, which takes into account events that have scattered in the shielding, but neglects any deflection from such scattering. For lower (higher) masses, we expect the thick orange (dashed green) line to more accurately describe the upper boundary of the excluded region.}
\label{ex-Nev-excl}
\end{center}
\end{figure*}

The rate for such scattering events can, of course, be suppressed if the dark matter is prevented from reaching the detector~\cite{Goodman:1984dc}. Such particles can scatter in the material above the detector, reducing the sensitivity of direct detection experiments to this class of dark matter candidates. While this effect has long been known, a simple analytic approach to the problem has hitherto been lacking. To obtain robust and conservative limits on the dark matter scattering cross section, we propose that the results of direct detection experiments be interpreted as follows. Let us define the published lower limit for a given experiment as $\sigma_{\rm low}$. In general, $\sigma_{\rm low}$ is obtained assuming that the dark matter reaches the detector with the standard velocity distribution and density.\footnote{The published lower bounds cited here each assume spin-independent scattering with $f_p=f_n$. We rescale as needed for different interactions and couplings.} If this is correct, then the number of expected events for any other cross section can be found by simply rescaling the lower bound, $N_{\rm events} \propto \sigma_{\psi p}/\sigma_{\rm low}$. For sufficiently large cross sections, however, the distribution of dark matter particles at the detector will not be the same as in the standard case, since particles may undergo one or more scatterings in the Earth or in its atmosphere. For very large values of $\sigma_{\psi p}$, the probability of scattering in the overburden can substantially reduce the rate of scattering events in the detector. To quantify this effect, we define the optical depth for scattering off of a target nucleus $A$ over a distance $r$ as $\tau_A =  \int dr \, n_A(r) \sigma_{\psi A}$. The expected number of events due to unscattered particles at the detector is then given as follows:
\beq \label{n_expec}
\frac{N_{\rm events}^{\rm unsc.}}{N_{\rm observed}} = \frac{\sigma_{\psi p}}{\sigma_{\rm low}} \, \exp\pL -   \sum_A \sigma_{\psi A} \int dr \, n_A(r) \pR,
\eeq
where we include the optical depth due to scattering off of all constituents of the overburden, including the atmosphere and the surface of the Earth as appropriate. We use the {\tt NRLMSISE-00} model \cite{NRLMSISE} for the Earth's atmosphere, and we assume that dark matter particles arrive from an angle of $54^\circ$ relative to directly overhead, which is the average angle with respect to the Earth's velocity vector at the relevant latitude \cite{Kavanagh:2017cru}. Because the optical depth is only a function of the location of the detector and does not depend on the dark matter velocity, it provides a uniform suppression for all dark matter particles from a given direction. Sensitive timing information on the events could potentially and the inclusion of an energy-dependent form factor could plausibly correct these statements by a factor of a few at most \cite{Kav-com}, probably in offsetting directions.

As is evident from \Eq{n_expec}, the behavior of $N_{\rm events}^{\rm unsc.}$ is non-monotonic as a function of $\sigma_{\psi p}$. The maximum of \Eq{n_expec} occurs for $\tau=1$, or at
\beq \label{sig-max}
\sigma_{\psi p}^{\rm max} = \cbL \sum_A \pL \frac{\mu_{\psi A}}{\mu_{\psi p}}\pR^2 \bL \cdots \bR  \int dr \, n_A(r) \cbR^{-1},
\eeq
where $\bL \cdots \bR$ represents the multiplicative factor to the right of the bracket in \Eq{sigT}. This factor is equal to $A^2$ for spin-independent scattering with $f_n=f_p$, $Z^2$ for spin-independent scattering with $f_n=0$, and to the appropriate spin values in the case of spin-dependent scattering. In the left panel of \Fig{ex-Nev-excl}, we plot $N_{\rm events}^{\rm unsc.}/N_{\rm observed}$ as a function of $\sigma_{\psi p}$ for $m_\psi =1\gev$ and for the conditions of the CRESST surface run \cite{Angloher:2017sxg}. In the right panel of \Fig{ex-Nev-excl} we plot the corresponding contours in the $m_\psi - \sigma_{\psi p}$ plane for the case of spin-independent couplings and $f_p=f_n$. We can conservatively and robustly rule out the range of cross sections between the two solid contours. If we less conservatively include events which scatter once or multiple times in the atmosphere, the upper boundary of the excluded region will fall somewhere between the thick orange and dashed green contours. For lower (higher) masses, we expect the thick orange (dashed green) line to more accurately describe the upper boundary of the excluded region.

\begin{figure*}[t]
\begin{center}
\includegraphics[width=\textwidth]{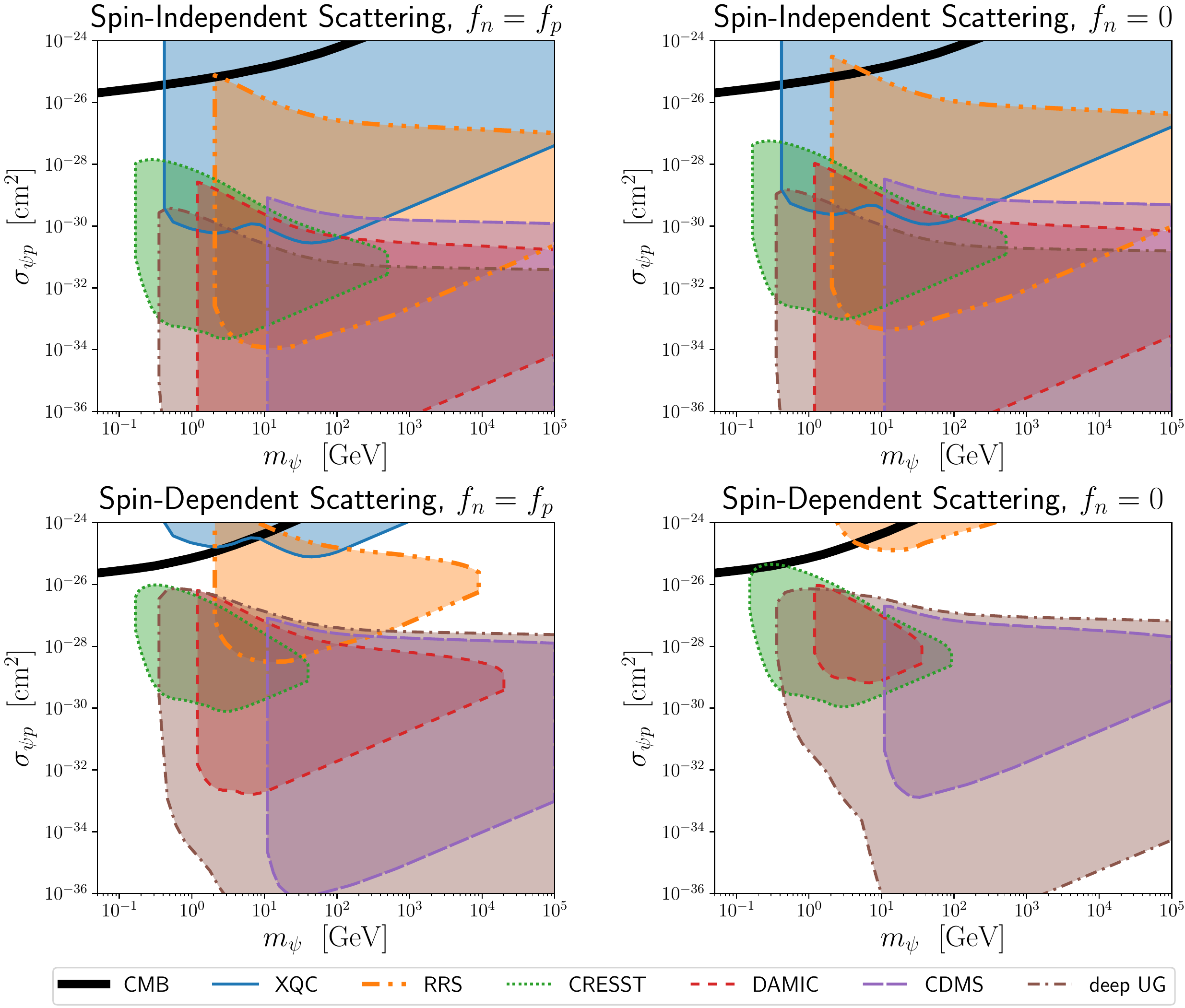} 
\caption{Robust and conservative bounds on the dark matter's scattering cross section with nucleons. The different panels correspond to spin-independent (top) and spin-dependent (bottom) couplings, and to the case of equal couplings to protons and neutrons (left) and to couplings only to protons (right). Inside of each of the shaded regions, the expected rate due to {\it unscattered dark matter particles alone} exceeds the rate at which the collaborations have reported limits. Including scattered events could increase the upper boundaries of some of these regions, in particular in the case of heavy dark matter candidates.}
\label{dd-bounds}
\end{center}
\end{figure*}

There are two features of \Eq{n_expec} that we would like to emphasize. First, there is no explicit velocity dependence in the optical depth, and thus no additional velocity dependence needs to be taken into account in recasting the limits. In other words, the velocity distribution of unscattered dark matter particles is not changed in our treatment. The change is only seen as a deficit in the overall number of dark matter particles at the detector. Second, the only dependence on $m_\psi$ is from the ratio $\pL \mu_{\psi A}/\mu_{\psi p}\pR^2$ in \Eq{sigT}. Thus isocontours of expected events in the $m_\psi -\sigma_{\psi p}$ plane will have a strong dependence on $m_\psi$ between $m_p$ and $m_A$, while at higher masses there is only a logarithmic dependence on $m_\psi$ inherited from $\sigma_{\rm low}$.


\begin{table*}[htp]
\begin{center}
\begin{tabular}{|c|c|c|c|c|c|c|}
\hline
experiment & element of interest & $A$ & $\%_i$ & $\vev{S_p}$ & $\vev{S_n}$ & $J_i$
\\ \hline \hline
\multirow{5}{*}{ XQC  \cite{McCammon:2002gb,Erickcek:2007jv}}
& (overburden${}^a$) hydrogen: $Z=1$ & 1 & $-$ & 1/2 & 0 & 1/2 \\  \cline{3-7}
& (overburden${}^a$) helium: $Z=2$ & 4 & $-$ & -- & -- & -- \\  \cline{2-7}
& \multirow{3}{*}{(detector) silicon: $Z=14$} & 28 & 92.2 & -- & -- & -- \\
&& 29 & 4.7 & $-0.002$ & $0.13$ & 1/2 \\
&& 30 & 3.1 & -- & -- & -- \\ \hline 
\multirow{6}{*}{ RRS  \cite{Rich:1987st, Starkman:1990nj}}
& \multirow{2}{*}{(overburden${}^b$) nitrogen: $Z=7$} & 14 & 99.6 & 1/2 & 1/2 & 1 \\ 
&& 15  & 0.4 & $-0.136$ & $0.028$  & 1/2  \\ \cline{3-7}
& \multirow{3}{*}{(overburden${}^b$) oxygen: $Z=8$} & 16 & 99.8 & -- & -- & -- \\
& & 17 & 0.04 & $-0.008$ & $0.48$ & 5/2 \\
& & 18 & 0.2 & -- & -- & -- \\ \cline{2-7}
& (detector) silicon: $Z=14$ & (see above) & \dots &   &   &  \\ \hline \hline
\multirow{3}{*}{ CRESST surface run  \cite{Angloher:2017sxg}}
& (overburden${}^c$) silicon: $Z=14$ & (see above) & \dots &   &   &   \\ \cline{2-7}
& (detector) aluminum: $Z=13$ & 27 & 100 & $ 0.326$ & $0.058$ & 5/2 \\ \cline{3-7}
& (detector) oxygen: $Z=8$ & (see above) & \dots &   &   & \\ \hline 
\multirow{6}{*}{ CDMS surface run  \cite{Abusaidi:2000wg, Abrams:2002nb}}
& (overburden${}^c$) silicon: $Z=14$ & (see above) & \dots &   &   &   \\ \cline{2-7}
& \multirow{5}{*}{(detector) germanium: $Z=32$}& 70 & 20.52 & -- & -- & -- \\ 
&& 72 & 27.45 & -- & -- & -- \\ 
&& 73 & 7.76 & $0.01$ & $0.42$ & 9/2 \\ 
&& 74 & 36.52 & -- & -- & -- \\ 
&& 76 & 7.75 & -- & -- & -- \\ \hline
\multirow{2}{*}{ DAMIC \cite{Aguilar-Arevalo:2016ndq}}
& (overburden${}^c$) silicon: $Z=14$ & (see above) & \dots &   &   &   \\ \cline{2-7}
& (detector) silicon: $Z=14$ & (see above) & \dots &   &   & \\ \hline
\multicolumn{7}{l}{${}^a$Galactic composition} \\ 
\multicolumn{7}{l}{${}^b$Atmospheric abundances (other elements included but subdominant)} \\
\multicolumn{7}{l}{${}^c$Using depths quoted in meters of water equivalent (mwe)} \\ 
\end{tabular}
\end{center}
\label{default}
\caption{Characteristics of the detectors and their overburden for the direct detection experiments discussed in this study. In the column $\%_i$ we list the percentage natural abundance of the given isotope. We use average values of the spin content per nucleon from Ref.~\cite{Ressell:1993qm} for ${}^{29}$Si and ${}^{73}$Ge and Ref.~\cite{Bednyakov:2004xq} for the remaining.}
\label{tableone}
\end{table*}%


Following this approach, we plot in \Fig{dd-bounds} conservative and robust constraints on the dark matter's scattering cross section with nucleons. This includes constraints from the XQC satellite experiment \cite{McCammon:2002gb, Erickcek:2007jv, Mahdawi:2017cxz}, the RRS balloon experiment (which we assume flew at 50 km to match their reported column depth) \cite{Rich:1987st}, the CRESST surface run (for which we assume only an atmospheric overburden) \cite{Angloher:2017sxg}, the CDMS surface run (for which we assume atmospheric and 10 meters of water equivalent rock overburden) \cite{Abusaidi:2000wg, Abrams:2002nb}, and the DAMIC shallow site run (for which we assume atmospheric and 100 meters water equivalent rock overburden) \cite{Aguilar-Arevalo:2016ndq}. Selected properties of these experiments and their shielding are listed in Table~\ref{tableone}. We also show in this figure a compilation of constraints from deep underground sites, including CRESST-III \cite{Petricca:2017zdp}, CDMSlite \cite{Agnese:2017jvy}, and modern xenon-based experiments \cite{Akerib:2016vxi, Aprile:2017iyp, Cui:2017nnn}, which we uniformly model with an overburden of the atmosphere plus 2000 meters water equivalent of rock, which is a mild underestimate of the true depth. Inside of each of these shaded regions, the expected rate due to {\it unscattered dark matter particles alone} exceeds the rate at which the collaborations have reported limits. Including scattered events could increase the upper boundaries of some of these regions, in particular for large masses.

In \Fig{dd-bounds}, we do not present any bounds which rely on dark matter particles reflected from the Sun \cite{An:2017ojc}, which require traveling through additional regions of high particle density. In addition, photon bremsstrahlung from scattering off of nuclei has been suggested as a novel detection mechanism in this range \cite{Kouvaris:2016afs}. We find, however, that unscattered dark matter particles are unable to induce these events. We also omit other space-based instruments that have the same detector target as XQC but higher thresholds, such as those discussed in Refs.~\cite{Starkman:1990nj, Mack:2007xj}. 

It is now appropriate to ask how reflective these constraints are of the actual bounds that would be derived after fully accounting for the dark matter particles deflected by shielding. Because the expected angular deflection and fractional change in momentum each scale with $\mu_{\psi A}/m_\psi$, low-mass dark matter particles are likely to be deflected by large angles when they scatter in the atmosphere or in other shielding, increasing their overall path length and preventing a large population of through-going dark matter particles from reaching the detector~\cite{Kavanagh:2017cru}. In principle, such particles could even thermalize with the Earth and become indistinguishable from thermal neutrons, and likely be removed from the signal analysis. The upper boundaries of our exclusion regions are thus a reasonable approximation to the truth in the case of $m_\psi \lesssim m_A$. For heavier particles, however, we expect a potentially non-negligible fraction of the scattered population to reach the detector with enough velocity to exceed the detection threshold~\cite{Emken:2017erx, Davis:2017noy, Kavanagh:2017cru}. For this reason, ruling out only the parameter space in which $N_{\rm events}^{\rm unsc.} \geq N_{\rm observed}$ necessarily constitutes a {\it conservative} bound, since the true bound may lie at somewhat larger cross sections where an admixture of scattered and unscattered dark matter particles are observed in the detector, especially in the case of large $m_\psi$. The kinetic energy of a very massive dark matter particle as a function of distance traveled can in principle be modeled as propagation through a medium that induces a continuous energy loss \cite{Kavanagh:2017cru}. If the path length is not extended by large-angle scatters, the final energy can be obtained numerically. Dedicated Monte Carlo is needed to understand the effects of the path length extension and to clarify the value of $m_\psi$ at which the cross-over between the low- and high-mass approximations occurs. In summary, we expect that the true limits should be close to our ``unscattered'' result (thick orange) at $m_\psi \lesssim m_A$ and close to the result of Ref.~\cite{Kavanagh:2017cru, Davis:2017noy} (dashed green) for $m_\psi \gg m_A$.

In the case of dark matter in the form of a stable six-quark dibaryon, one would naively expect such a particle to have a large spin-independent cross section with nucleons, which would appear to be ruled out by the above analysis. To evade this conclusion would require a very strong suppression of the scattering rate. We also note that cross sections of the size discussed here have been suggested to significantly alter the dark matter's local velocity distribution~\cite{Mahdawi:2017cxz}. As a rough estimate, a dark matter particle will scatter with gas at a rate of $\sim 10^{-2} \,{\rm Gyr}^{-1} \, (\sigma_{\psi p}/10^{-26} \,{\rm cm}^2)(n_{\rm gas}/{\rm cm}^{-3})$. Although the consequences of this effect have not been worked out in detail, this estimate leads us to conclude that our results should not be substantially impacted by such interactions.


\section{Gamma-Ray Signatures of Cosmic-Ray Interactions With Dark Matter}
\label{gamma-ray-sec}

The interactions of cosmic-ray protons with dark matter could potentially generate an observable flux of gamma rays within this class of models. This process is analogous to the standard production of gamma rays through the decays of neutral pions generated in the collisions of cosmic-ray protons with interstellar gas, but with the dark matter in this case playing the role of the target gas. 

With this possibility in mind, previous studies have derived constraints on such models based on gamma-ray observations of the inner Milky Way~\cite{Cyburt:2002uw, Mack:2012ju}. For two reasons, however, it is difficult to use gamma-ray observations of the region surrounding the Galactic Center to derive robust constraints on the scattering cross section for cosmic rays with dark matter. First, given the large scattering cross sections in this class of models, it is unclear what density profile we should expect the dark matter to be described by, especially in the inner regions of the Milky Way's halo~\cite{Spergel:1999mh,Rocha:2012jg,Tulin:2013teo,Dave:2000ar}. In particular, such interactions could be expected to lead to the destruction of any central density cusp. Second, the number density of gas targets significantly exceeds that of dark matter particles in the Inner Galaxy, making any contribution to the gamma-ray flux from cosmic-ray scattering with dark matter likely to be highly subdominant compared to the emission from conventional processes. As a consequence, it would be difficult to use existing observations to either identify or robustly exclude the presence of a gamma-ray signal arising from cosmic-ray scattering with dark matter in the Inner Galaxy. 

For these reasons, we argue here that a more promising target for gamma-ray telescopes in this class of models is the high-latitude sky. Although cosmic-ray protons are generated by sources ({\it i.e.}~supernova remnants) that are distributed throughout the disk of the Milky Way, these particles undergo diffusion and escape the disk on a timescale of $t_{\rm esc} \sim z_s^2/4 D(E_p) \sim 10^{8} \,{\rm yr} \times (z_s/4\,{\rm kpc})^2 \, (E_p/{\rm GeV})^{-1/3}$, where we have adopted a diffusion coefficient of $D \approx 1.5 \times 10^{28} \, (E_p/{\rm GeV})^{1/3}$ cm$^2$/s throughout the Galactic Disk, in agreement with boron-to-carbon and other local cosmic-ray measurements~\cite{Johannesson:2016rlh,Porter:2017vaa,Benyamin:2016xcq,Simet:2009ne,Hooper:2017tkg}. After escaping the disk, these cosmic rays then go on to diffuse more rapidly throughout the bulk of the Milky Way's extended halo. While in the extended halo, cosmic-ray scattering with dark matter could plausibly compete with, or even dominate over, interactions with gas. Furthermore, the dark matter distribution within the extended halo of the Milky Way is expected to be robust, since the densities are low enough that it is unlikely that a single dark matter particle has been significantly impacted by repeated scattering, either with other dark matter or baryons.

\begin{figure*}[t]
\begin{center}
\includegraphics[width=0.47\textwidth]{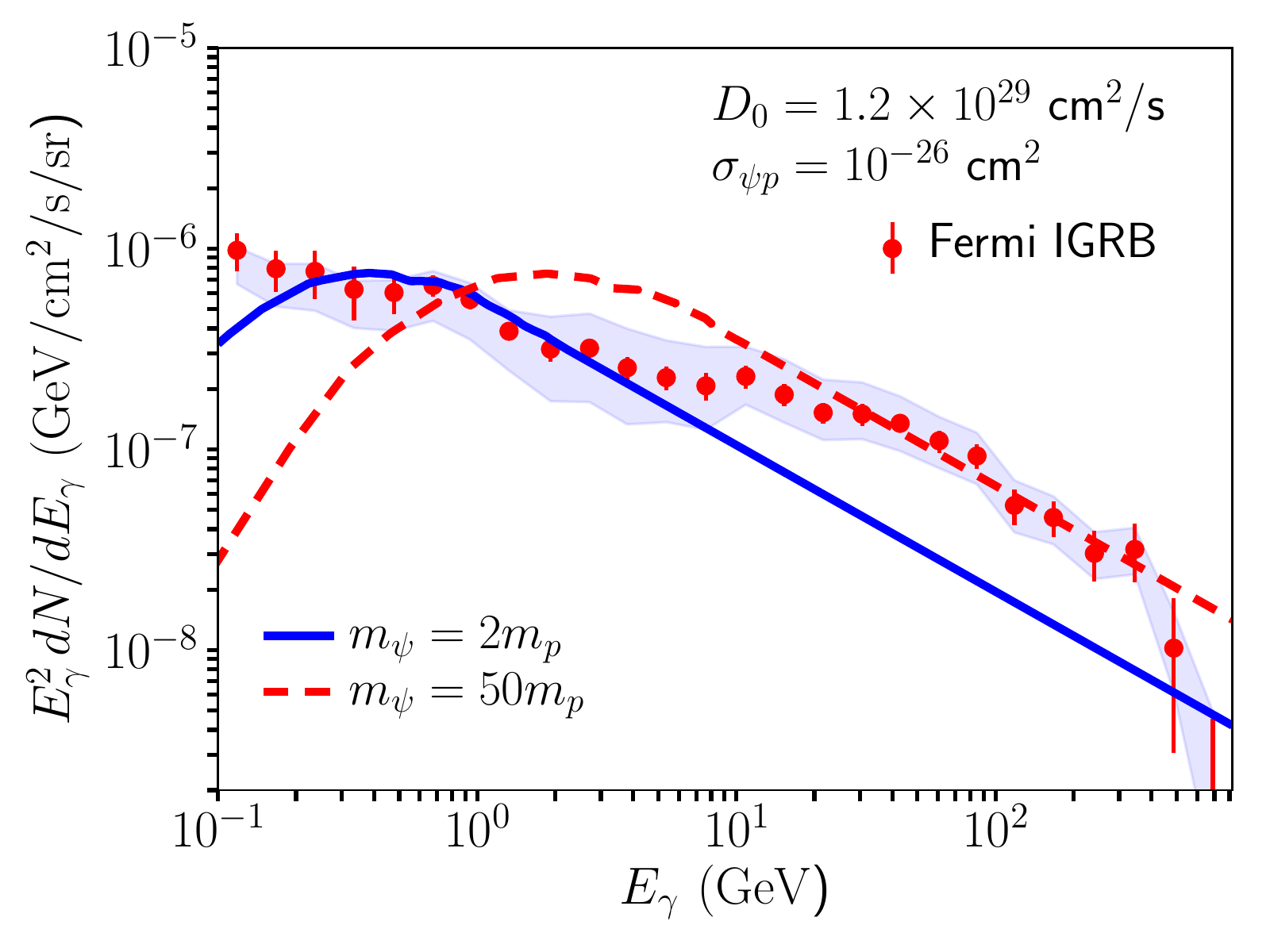}
\includegraphics[width=0.47\textwidth]{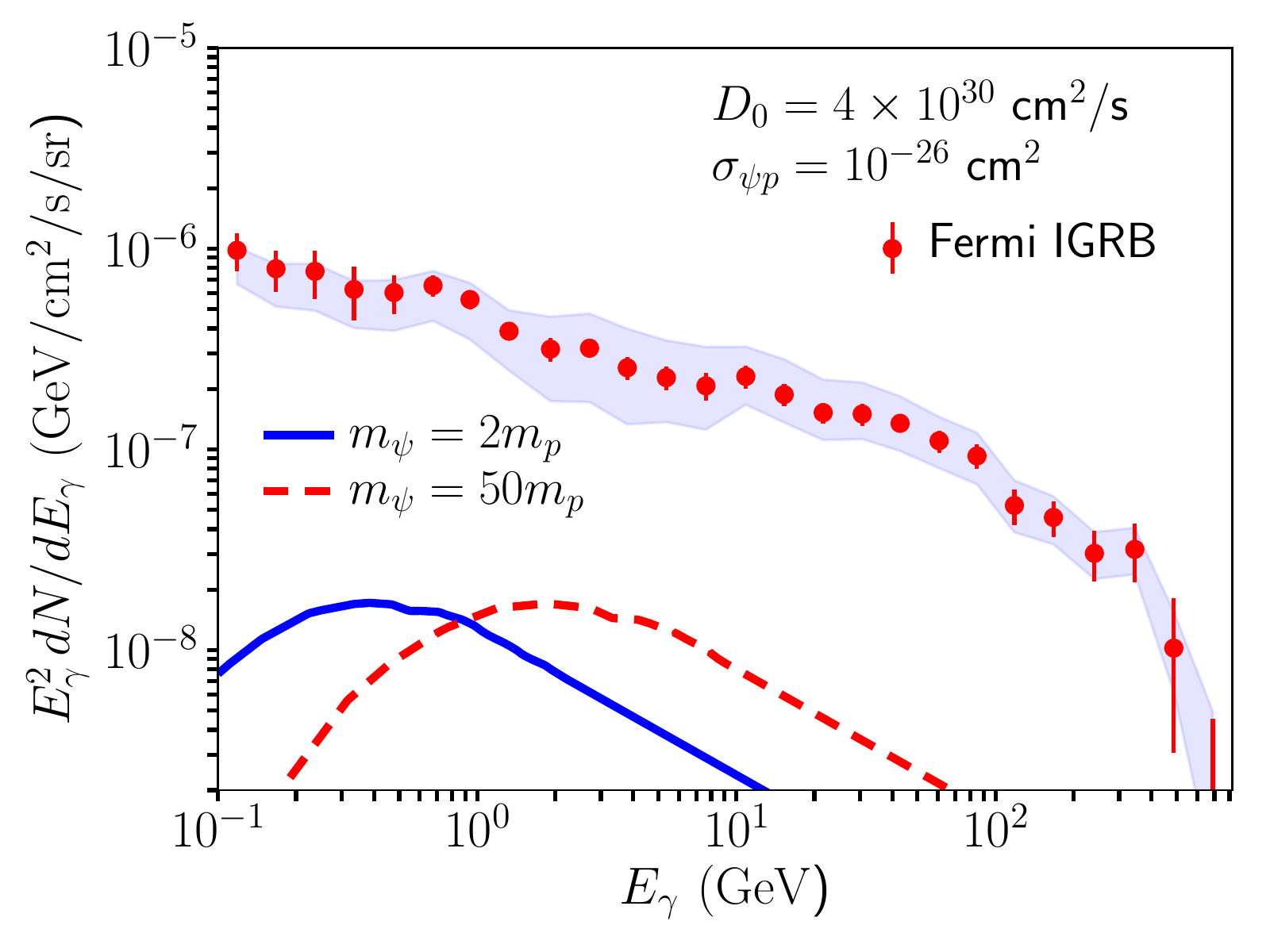}
\caption{The gamma-ray spectrum from cosmic-ray scattering with dark matter in the extended halo of the Milky Way, averaged over the high-latitude ($|b| > 30^{\circ}$) sky, compared to the isotropic gamma-ray background (IGRB) as reported by the Fermi Collaboration. Results are shown for two values of the halo diffusion coefficient and of the dark matter mass. Although the relevant uncertainties are significant, it is possible that dark matter models featuring large cross sections with protons could result in potentially observable contributions to the IGRB.}
\label{fig:gamma}
\end{center}
\end{figure*}

Although the gamma-ray flux that is observed at high galactic latitudes is often referred to as the extragalactic gamma-ray background (EGB), a portion of this emission could arise from processes taking place within the halo of the Milky Way. This has been considered previously within the context of cosmic-ray interactions with either circum-galactic gas~\cite{Feldmann:2012rx} or radiation~\cite{Keshet:2003xc}. The component of the EGB that has not been resolved into emission from individual sources or attributed to Galactic diffuse emission processes is known as the isotropic gamma-ray background (IGRB), and this has been measured by the Fermi Collaboration over energies between 0.1 and 820 GeV~\cite{Ackermann:2014usa}. Although the detailed origin of this emission is still being debated, there is considerable empirical support for a scenario in which both non-blazar active galaxies~\cite{Hooper:2016gjy} and star-forming galaxies~\cite{Linden:2016fdd} (see also Refs.~\cite{Tamborra:2014xia,Ackermann:2012vca,Stecker:2010di,DiMauro:2013xta}) provide the largest contributions, along with smaller but potentially non-negligible contributions from blazars~\cite{Cuoco:2012yf,Harding:2012gk,Ajello:2011zi,Ajello:2013lka,Stecker:2010di}, galaxy clusters~\cite{Zandanel:2014pva}, propagating ultra-high energy cosmic rays~\cite{Taylor:2015rla,Ahlers:2011sd}, and perhaps even annihilating dark matter particles~\cite{Ackermann:2015tah,DiMauro:2015tfa,Ajello:2015mfa,Cholis:2013ena}. More quantitatively, Ref.~\cite{Hooper:2016gjy} concludes that unresolved non-blazar active galaxies account for no less than 59\% of the IGRB photons above 1 GeV, while Ref.~\cite{Linden:2016fdd} finds that star-forming galaxies are responsible for at least 24\% of the IGRB intensity above 1 GeV (each at the 2$\sigma$ confidence level). This class of scenarios is further supported by the results of previous analyses~\cite{Fornasa:2015qua,DiMauro:2016cbj,DiMauro:2015tfa,Ajello:2015mfa,Cholis:2013ena,Cavadini:2011ig,Siegal-Gaskins:2013tga}, including cross-correlation studies of the IGRB with multi-wavelength data~\cite{Xia:2015wka,Cuoco:2015rfa,Shirasaki:2014noa,Shirasaki:2015nqp,Ando:2015bva}. In light of this, we consider any scenario involving cosmic-ray interactions with dark matter to be ruled out if it leads to a flux of gamma rays at high latitudes that is larger than 20\% of the measured IGRB at a given energy. On the other hand, a somewhat smaller contribution of this type could potentially be identified within the Fermi dataset, or within the data of next-generation space-based gamma-ray telescopes.

To provide a quantitive estimate for the gamma-ray signal generated in the interactions of cosmic rays with dark matter, we first calculate the distribution of cosmic-ray protons in the extended halo of the Milky Way. To this end, we solve the following steady-state diffusion equation:
\begin{equation}
0 = \vec{\nabla} \cdot [ D(E_p) \vec{\nabla} \frac{dN_p}{dE_p} (\vec{x},t,E_p) ] +Q(\vec{x},t,E_p),
\end{equation}
where $D$ is the diffusion coefficient for the extended halo of the Milky Way (as opposed to within the region surrounding the Galactic Disk), and $dN_p/dE_p$ is the distribution of cosmic rays. Although the value of the halo's diffusion coefficient is quite uncertain, we follow Ref.~\cite{Feldmann:2012rx} in adopting two benchmarks intended to roughly bracket a plausible range of values: $D_0=1.2\times 10^{29}$ cm$^2$/s and $D_0=4\times 10^{30}$ cm$^2$/s, where $D=D_0 \times (E_p/{\rm GeV})^{1/3}$. For comparison, $D_0 \approx 1.5 \times 10^{28}$ cm$^2/$s within and near the disk of the Milky Way.

The quantity $Q$ is the source term, which describes the spectrum and spatial distribution of the cosmic rays injected from the disk into the surrounding halo. For the spatial distribution of cosmic-ray sources along the Galactic Plane, we adopt a Lorimer profile~\cite{Lorimer:2003qc}:
\begin{equation}
Q(R,t,E_p) = Q_0\, E_p^{-2.4} \, R^{2.35}\, \exp(-R/1.528\,{\rm kpc}) \,f(t),
\end{equation}
where $R$ is the distance from the Galactic Center and $Q_0$ is a normalization constant chosen to reproduce the local cosmic-ray spectrum. For the time dependence of cosmic-ray injection, we adopt a profile based on the estimated star formation history of the Milky Way: $f(t)=1+t/(1\, {\rm Gyr})$ for $t\le 2$ Gyr, $f(t)=3$ for 2 Gyr $< t \le 6$ Gyr and $f(t)=3-0.5(t-6\,{\rm Gyr})$ for 6 Gyr $< t \le 10$ Gyr.

\begin{figure*}[t]
\begin{center}
\includegraphics[width=\textwidth]{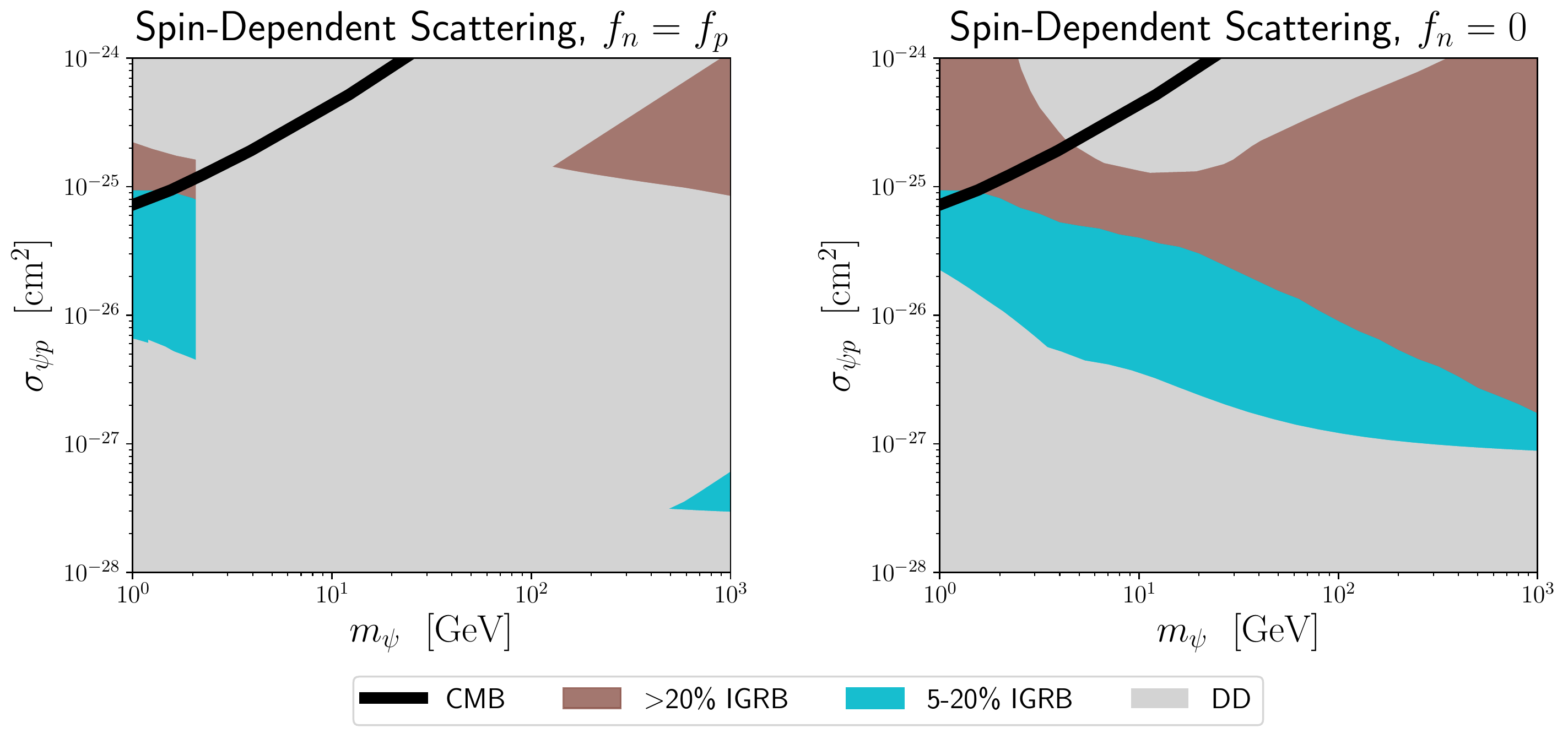}
\caption{The bright blue band represents the range of parameter space in which between 5\% and 20\% of the measured isotropic gamma-ray background (IGRB) may be generated by cosmic-ray interactions with dark matter for values of the halo diffusion coefficient between $D_0=(1.2-40)\times 10^{-29}$ cm$^2$/s. In this range, it is possible that cosmic-ray interactions with dark matter generate a non-negligible and potentially observable fraction of the IGRB. Parameter space above this band is likely ruled out. We compare this to constraints from direct detection experiments (shaded grey), and from the cosmic microwave background~\cite{Gluscevic:2017ywp} (thick solid line).}
\label{constraints2}
\end{center}
\end{figure*}

Once in the halo, we calculate the spectrum of gamma rays that these cosmic rays produce through scattering with dark matter. The spectrum of neutral pions generated through these interactions can be written as follows:
\begin{equation}
\frac{dN_{\pi}}{dE_{\pi}} = 4\pi n_{\rm \psi} \int^{\infty}_{E^{\rm min}_{p}(E_{\pi})}  dE_{p} J_p(E_p) \frac{d\sigma_{\pi}}{dE_{\pi}}(E_{\pi}, E_p, m_{\psi}),
\end{equation}
where $J_p$ is the intensity of cosmic-ray protons and $n_{\psi}$ is the number density of dark matter particles, which we take to be described by an NFW profile with a scale radius of 20 kpc and normalized to a local density of 0.4 GeV/cm$^3$. The quantity $d\sigma_{\pi}/dE_{\pi}$ is differential cross section for the production of neutral pions. We treat the normalization of this cross section as a free parameter, and adopt a spectral shape for the pions following the parameterization described in Ref.~\cite{Blattnig:2000zf}, adjusted to account for the mass of the dark matter candidate. From this spectrum of pions, we can calculate the spectrum of gamma rays that is produced in their decays:
\begin{equation}
\frac{dN_{\gamma}}{dE_{\gamma}} = 2 \int^{\infty}_{E^{\rm min}_{\pi}(E_{\gamma})}  dE_{\pi} \frac{dN_{\pi}}{dE_{\pi}} \frac{1}{\sqrt{E^2_{\pi}-m^2_{\pi}}}.
\end{equation}
Lastly, we integrate the above expressions over the line-of-sight to obtain a differential spectrum of gamma rays generated through cosmic-ray scattering with dark matter, as a function of the direction observed. To distinguish the signal in question from that generated in the disk of the Milky, we perform this integral only beyond 4 kpc from the Galactic Disk, and thus our results conservatively underestimate the total expected signal.

In Fig.~\ref{fig:gamma}, we show the contribution to the IGRB from cosmic-ray scattering with dark matter, for two choices of the halo diffusion coefficient and for two values of the dark matter mass. For the smaller value of $D$ adopted in the left frame, the cosmic-ray halo is largely concentrated within the innermost few tens of kpc, where dark matter particles are abundant (being within or near the scale radius of the Milky Way's dark matter profile), thus generating a substantial fraction of the IGRB. For the larger diffusion coefficient adopted in the right frame, the cosmic-ray halo is more extended, suppressing the overall gamma-ray emission from interactions with dark matter. 

There are at least two ways in which a high-latitude component of gamma rays from cosmic-ray scattering with dark matter could potentially be distinguished from the remainder of the IGRB. Firstly, the spectrum arising from such interactions includes a feature that is similar to that from cosmic-ray scattering with gas, but with a spectral shape that is determined by the mass of the dark matter. Secondly, we point out that this gamma-ray signal is not strictly isotropic, unlike contributions from unresolved cosmological source populations. More specifically, we find that the gamma-ray flux from these interactions is approximately a factor of 3 higher at $(l,b)=(0^{\circ},30^{\circ})$ than at $b=90^{\circ}$, and a factor of 3 lower at $(l,b)=(180^{\circ},30^{\circ})$. This gradient could be used to identify a subdominant signal from cosmic-ray interactions in the halo, either with dark matter or with circum-galactic gas.

In Fig.~\ref{constraints2}, we plot the range of parameter space for which, at the peak of the spectral feature, between 5\% and 20\% of the measured IGRB may be generated by cosmic-ray interactions with dark matter. In calculating this band, we have marginalized over the range of values of the halo diffusion coefficient, as considered in Fig.~\ref{fig:gamma}. In the parameter space above this band, we expect such interactions to generate more than 20\% of the observed IGRB (at some energy), in considerable tension with the evidence that the IGRB is dominated by contributions from non-blazar active galaxies and starforming galaxies~\cite{Hooper:2016gjy,Linden:2016fdd}. 

In the region of parameter space with $\sigma_{\psi p} \sim 10^{-26} \, {\rm cm}^2$, a non-negligible and potentially observable component of the IGRB could originate from cosmic-ray interactions with dark matter, while not being excluded by other considerations.

\section{Conclusions}

We find that there exists a range of potentially viable parameter space in which the dark matter possesses a very large cross section with nucleons. In revisiting this class of models, we confirm bounds due to the disassociation of nuclei during Big Bang nucleosynthesis (BBN). We have also presented a simple, analytic method for obtaining conservative but robust limits from direct detection experiments, written succinctly in \Eq{n_expec}. Using this method, we calculate excluded regions of parameter space in which the number of events from dark matter particles that scatter in the detector, and not in the overburden, exceeds the observed number of events. For cross sections slightly above our excluded zones, Monte Carlo simulations are necessary to understand the recoil spectrum from a dark matter particle with given parameters. Although we do not find any viable parameter space in which the dark matter has a large spin-independent cross section with nucleons, viable spin-dependent parameter space does exist, in particular near $m_\psi \sim \gev$ and $\sigma_{\psi p} \sim 10^{-26} \cm^2$ or for much higher dark matter masses (see \Fig{dd-bounds}).

We have also revisited the gamma-ray signatures predicted in this class of models. In the range of parameter space allowed by direct detection experiments and other constraints, cosmic-ray scattering with dark matter in the extended halo of the Milky Way could generate a non-negligible fraction of the diffuse gamma-ray emission observed at high galactic latitudes. This contribution would constitute a novel and distinctive signature, potentially observable by Fermi or future space-based gamma-ray telescopes. 


\begin{acknowledgments}  
We would like to thank Adrienne Erickcek, Brian Fields, Paddy Fox and Vera Gluscevic for helpful communication. SDM would also like to acknowledge Prateek Agrawal and David Pinner for particularly valuable discussions. We acknowledge a very helpful exchange with Bradley Kavanagh that clarified the effects of (and uncertainties due to) the use of an averaged incoming dark matter velocity vector and the omission of a scattering form factor. This manuscript has been authored by Fermi Research Alliance, LLC under Contract No. DE-AC02-07CH11359 with the U.S. Department of Energy, Office of Science, Office of High Energy Physics. The United States Government retains and the publisher, by accepting the article for publication, acknowledges that the United States Government retains a non-exclusive, paid-up, irrevocable, world-wide license to publish or reproduce the published form of this manuscript, or allow others to do so, for United States Government purposes.
\end{acknowledgments}
\bibliography{v-strong-b}

\end{document}